\documentclass{amsart}

\usepackage{setspace}

\usepackage{graphicx} 
\usepackage{array} 
\usepackage{bm} 
\usepackage{booktabs} 
\usepackage{ctable} 
\usepackage{subfig} 
\usepackage{caption} 
\usepackage{eurosym}
\usepackage{footnote}
\usepackage{threeparttable}
\usepackage{mathtools}
\usepackage{color}
\usepackage{rotating}
\usepackage{lscape}
\usepackage{tensor}
\usepackage[linewidth=1.5pt]{mdframed}
\usepackage{xcolor,colortbl}
\usepackage{pifont} 
\usepackage[export]{adjustbox}

\usepackage[round]{natbib}
\makeatletter 
\renewcommand\@biblabel[1]{#1.} 
\makeatother %

\definecolor{Gray}{gray}{0.85}

\numberwithin{equation}{section}

\begin{document}

\title[Bayesian binary quantile regression for Bachelor-Master transition]
{Bayesian binary quantile regression for the analysis of Bachelor-Master transition}

\author
[C.       Mollica]
{Cristina Mollica*}
\address
{*Dipartimento di Scienze statistiche\\
 Sapienza Universit\`a di Roma\\
 Piazzale A. Moro 5\\00185~Roma\\Italy}
\email{cristina.mollica@uniroma1.it}
\author
[L.   Petrella]
{Lea Petrella**}
\address
{**Dipartimento di Metodi e Modelli per l'Economia, il Territorio e la Finanza\\
 Sapienza Universit\`a di Roma\\
 Via del Castro Laurenziano, 9\\00161~Roma\\Italy}
\email{lea.petrella@uniroma1.it}
\thanks{Correspondence to: cristina.mollica@uniroma1.it}

\keywords{Binary quantile regression, Asymmetric Laplace distribution, Data augmentation, Gibbs sampling, Bachelor-Master transition, university drop-out}


\begin{abstract}
The multi-cycle organization of modern university systems stimulates the interest in studying the progression to higher level degree courses during the academic career. In particular, after the achievement of the first level qualification (Bachelor degree), students have to decide whether to continue their university studies, by enrolling in a second level (Master) programme, or to conclude their training experience.
In this work we propose a binary quantile regression approach to analyze the Bachelor-Master transition phenomenon with the adoption of the Bayesian inferential perspective. 
In addition to the traditional predictors of academic outcomes, such as the personal characteristics and the field of study, different aspects of the student's performance are considered. Moreover, a new contextual variable, indicating the type of university regulations, is taken into account in the model specification.
The utility of the Bayesian binary quantile regression to characterize the non-continuation decision after the first cycle studies is illustrated with an application to administrative data of Bachelor graduates at the School of Economics of Sapienza University of Rome and compared with a more conventional logistic regression approach.
\end{abstract}
\maketitle
%
\makeatletter \@setaddresses \makeatother \renewcommand{\addresses}{}


\section{Introduction}
\label{s:intro}
The set of regulations launched by the Bologna Process in 1999 deeply reformed the organization of the higher education systems in Europe, with the aim at expanding the participation to tertiary education and improving its outcomes.

In accordance with the reforms of the Bologna Declaration, the Italian university system is currently structured in three hierarchical levels of academic education: the 3-year first cycle degree (\textit{Bachelor degree}), the 2-year second cycle degree (\textit{Master degree}) and the 3-year third cycle research doctorate.
The multi-cycle organization of modern higher education systems offers the possibility to examine new academic outcomes of students' training path and also to assess well-studied phenomena in new critical steps of the university career, such as the delicate phase of transition between consecutive levels of the academic studies. The progression from the first to the second cycle studies, referred to as \textit{Bachelor-Master transition}, is a crucial step of the university experience, but its characterization still remains an open problem from both an institutional and a statistical perspective.
When considering the entire degree path, involving both the first and the second level qualification, the non-transition at Master level determines a specific type of students' non-completion behavior.
In particular, after the achievement of the Bachelor degree, students have to decide whether to continue their university studies, by enrolling in a second cycle programme, or to stop their training experience.
From this perspective, the decision to interrupt the academic career after the completion of the first cycle course leads to the drop-out of the university, that is, to the withdrawal of the Bachelor graduate (also known as \textit{undergraduate}) from the institution. 
In this paper we focus our attention on the non-progression decision after the first level qualification attainment and investigate the factors that can influence the choice of not embarking on a second level specializing degree course within the same institution of enrollment. 
Besides the classical predictors of academic outcomes, such as the personal characteristics and the field of study, a special attention is turned to different aspects of student's performance and on a contextual variable, indicating the type of academic regulations 
that the students experienced during their first level career. Investigating the factors that may influence the non-completion phenomenon permits universities to allocate the resources aimed at supporting the transition at Master level in a more efficient way. 

From a statistical point of view, the university drop-out analysis has been traditionally addressed with regression methods within the generalized linear model family (GLM) and extensions thereof,
by either adopting a binary definition of the response variable to discriminate between drop-out and continuation (see for example \cite{Dipietro-Cutillo} and \cite{Belloc2010} and references therein) or considering a more complex setting with multiple alternatives \citep{Belloc2011}. 
In this paper we consider a binary quantile regression (BQR) approach for the analysis of the non-continuation outcomes with the adoption of the Bayesian inferential framework. 
Bayesian methods are very useful and flexible tools accounting for parameter uncertainty by combining data with prior pre-experimental information.

Quantile regression (QR) provides a very useful device to explore as different location measures of the response distribution are affected by the predictors, in order to gain a more in-depth understanding of the relation between the outcome of interest and the explanatory variables. 
In so doing, it is possible to capture also ``extreme'' behaviors, not detectable by those models that take into account only the mean of the response variable.
The Bayesian QR framework usually relies the inference on the Asymmetric Laplace distribution (ALD), as described in the seminal works by \cite{Yu:Moyeed} and \cite{Kottas2001} and in subsequent contributions, such as \cite{Lum} and \cite{Bernardi2015}.
By using a data augmentation method, more recently QR modeling has been extended for the treatment of binary response variables, see \cite{Benoit:Van} and \cite{Benoit2013}.
Here we illustrate the utility of the Bayesian BQR approach to describe the non-continuation decision with an application to administrative data on students who attained their first level qualification at the School of Economics of Sapienza University of Rome during the period from the academic year (a.y.) 2009-2010 to 2012-2013. Up to our knowledge this represents the first attempt to investigate the Bachelor-Master transition phenomenon by means of a Bayesian binary regression model.

The remainder of the paper proceeds as follows. In Section~\ref{s:reform} we provide an outline of the higher education reform launched in 1999 and describe its effects on the current Italian university system. Section~\ref{s:overview} overviews the existing analyses of the Bachelor-Master transition, whereas Section~\ref{s:modeling} contains a review of the BQR model with the related MCMC methods to accomplish for a fully Bayesian estimation. The administrative data set and the results of the Bayesian BQR are presented in Section~\ref{s:empana}. The paper ends with concluding remarks and proposals of future developments in Section~\ref{s:conc}.

\section{The university reform in Italy}
\label{s:reform}
The Bologna Declaration in 1999 
marked the beginning of a radical and ambitious reform process of higher education in Europe, known as \textit{Bologna Process}.  
With the aim at creating an European Area of Higher Education and promoting its international competitiveness,
the major novelties introduced by the Bologna Process include: (i) the harmonization of the European higher education qualification systems, (ii) the adoption of tertiary education study programmes based on a multi-cycle structure and (iii) the introduction of a common system of credits to facilitate qualification comparisons and encourage students' mobility.

In Italy the adaptation to the objectives of the Bologna Process started with the Ministerial Decree (MD) 509/99, whose reforms became effective in the a.y. 2001-2002 and continued with the subsequent MD 270/04, that was applied starting from the a.y. 2008-2009.
The reforms led to a substantial remodeling of the architecture of the Italian university system. 
The fundamental change concerns the replacement of the single-cycle 4/5-year degree programme with a study path organized in three hierarchical levels of academic education: 
\begin{itemize}
\item[-]the 3-year first cycle degree (equivalent to the Bachelor degree), that provides both the basic knowledge 
and the necessary tools for the acquisition of adequate professional competence, in order to facilitate the entry in the world of work;
\item[-]the 2-year second cycle degree (equivalent to the Master degree), that offers an advanced level of training to prepare students for the exercise of specific professions or highly qualified activities requiring specific skills;
\item[-]the 3-year third cycle research doctorate, that represents the maximum level of academic education and the key step to take up a career based on the research activity.
\end{itemize}
Due to the duration of the first two academic levels, in Italy the reform is also referred to as \textit{``3+2'' degree system}.

As mentioned above, one of the most important reforms established by the Bologna Process is the introduction of the European Credit Transfer and Accumulation System (ECTS), that supplies a standardized criterion to describe higher education qualifications in the European Union through the adoption of a common measure of the study commitment needed for their achievement. 
The ECTS aims at increasing transparency of qualification comparisons and at facilitating students' mobility among the European education institutions thanks to the transfer of the ECTS credits (hereinafter, simply referred to as credits).
Students can progressively accumulate credits by passing the exams scheduled by their study plan.  
Each exam, in fact, is associated with a certain number of credits commensurate with the student workload
required to achieve the expected learning outcomes of the course.

The latter MD 270/04 brought a series of modifications and integrations to the former MD 509/99 with the declared aim at increasing the autonomy of the single academic institutions, mostly in the organization of their training services. 
Apart from some formal changes regarding the designation of the degree qualifications and of the groupings of the study courses into homogeneous degree classes, two important points 
deserve to be remarked for their potential impact on students' choice to start a second cycle programme.
The former entails the specification of a maximum number of exams for the qualification achievement, in order to assist a more coherent attribution of the credits with respect to the required workload and to avoid an extreme fragmentation of the training activities. This limit varies according to the qualification level and is equal, respectively, to 20 exams for the Bachelor degree and 12 exams for the Master degree.
The latter point concerns the set of reforms that regulate the access to a Master degree course and establishes a stronger separation between the first and the second cycle career. 
In particular, although the Bachelor degree is still a fundamental requisite, with the MD 270/04 the admission to a second level programme is constrained to the possession of curricular requirements and of an adequate personal preparation described by the academic regulations of the specific Master course and, hence, not necessarily dictated by the first level study path. This aspect facilitates undergraduates' mobility, in the sense that they have a major chance to start a Master level experience in a school different from that where they have earned the Bachelor degree.  
Moreover, unlike the MD 509/99, with the new MD 270/04 the exam grades of the first level career do not contribute to determine the final grade of the Master degree.
This set of reforms makes the prerequisites for accessing to a second cycle programme less restrictive and dependent on the previous first cycle studies.
The greater flexibility and the more careful recalibration of the number of activities in the study programme introduced by the MD 270/04 could represent an incentive for the undergraduates to continue their academic studies at Master level.
For this reason, in addition to other explanatory variables, we will analyze 
the effect of the MD, that is, the role of the academic regulation context that the students faced during the first level career on their continuation decisions.

\section{An overview of the Bachelor-Master transition literature}
\label{s:overview}

Whereas the interruption of attendance by enrolled students, called \textit{university drop-out} and examined in several works such as \cite{Smith-Naylor}, \cite{Dipietro-Cutillo}, \cite{Belloc2010}, \cite{Belloc2011}, \cite{Aina2013} and  \cite{Larsen2013a}, as well as the transition from university to work, see \cite{Biggeri}, \cite{Grilli2006}, \cite{Ballarino} and \cite{Pozzoli},
have been widely investigated in the higher education literature, the progression from the first to the second cycle academic studies has received much less attention. 
Nevertheless, the athenaeum is strongly interested in promoting and encouraging the completion of the ``3+2'' degree programme, in order to train highly qualified professional figures, who can better face the selection of the labour market, and to increase its competitiveness.
Moreover, an in-depth understanding of the mechanism of Bachelor-Master transition is also useful for the authorities to plan the training offer and optimize resource allocation.

\begin{figure}[t]
\centering
\includegraphics[scale=.45]{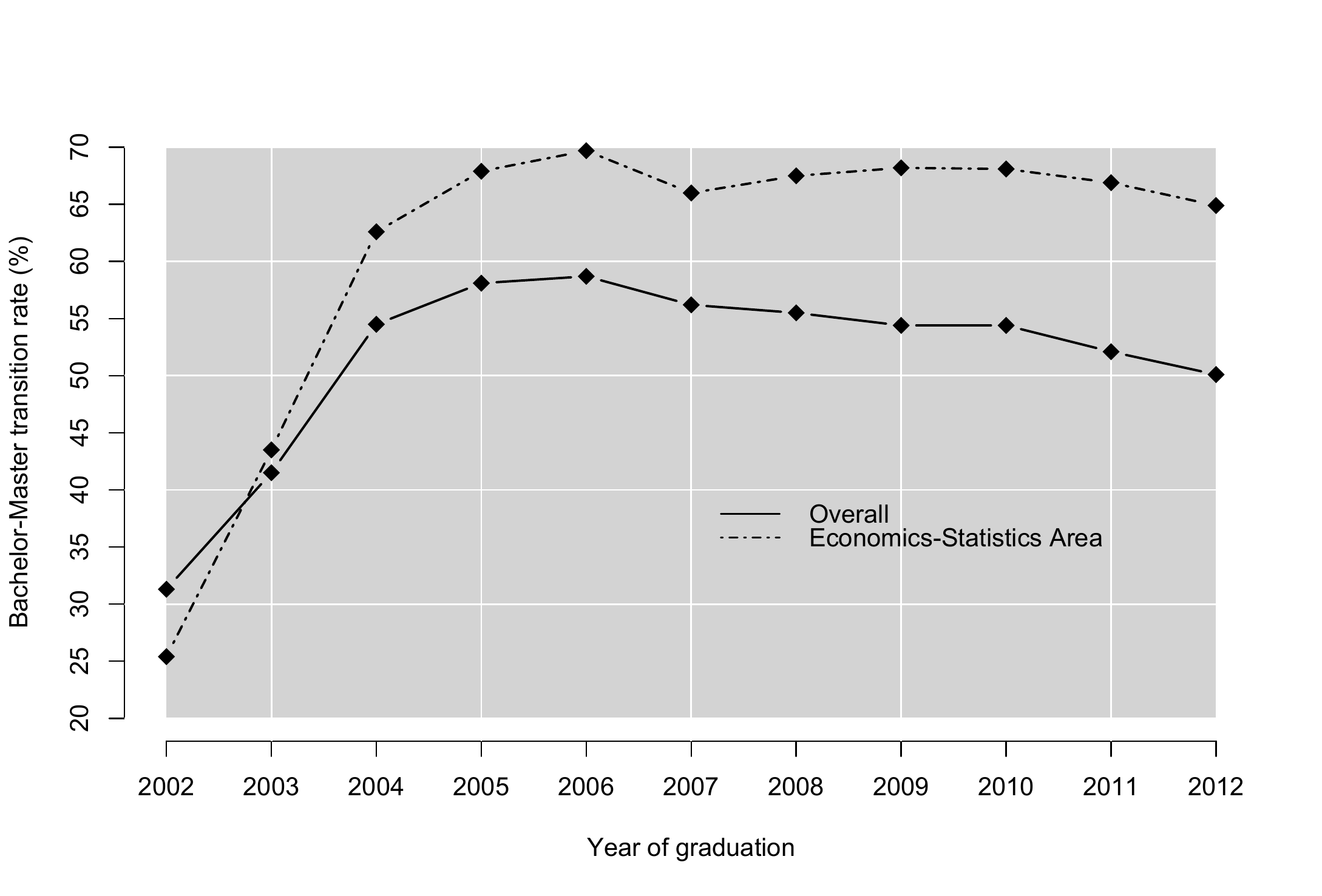}
\caption{Bachelor-Master transition rate (\%) in Italy: overall percentage (solid line) and the one for the Economics-Statistics area of study (dashed line). Source: \cite{ANVUR2014}.}
\label{fig:TimeSeriesTransition}
\end{figure}

From an institutional point of view the attention on the progression to the second cycle studies of the Italian Bachelor graduates is proved by the monitoring of the phenomenon by the National Agency for the Evaluation of Universities and Research Institutes (ANVUR) of the Italian Ministry of Education, University and Research. 
Figure~\ref{fig:TimeSeriesTransition} shows the trend of the Bachelor-Master transition rate for the decade 2002-2012, as reported in \cite{ANVUR2014}, where the solid line indicates the national average and the dashed one is specific for the Economics-Statistics area of study.
As remarked in the report, starting from 2005-2006 data are more stable and reliable, since in those years the ``3+2'' system became fully operating and the first genuine Bachelor cohorts faced with the transition decision. 
The proportion of undergraduates who progress to a Master degree course reached around 59\% in 2006, but slowly drops in the second part of the period and shrinks up to 50\% in 2012.
The continuation rate for the Economics-Statistics area is manifestly above the national values, ranging between 65-70\%, and follows an analogous decreasing tendency at the end of the period.
For organizational purposes, identifying the determinants of the decision to enroll in a Master degree course can concretely suggest to the academic institution how to incentivize and support the progression to higher level studies. 

Indeed, a factual research approach for the analysis of the transition between consecutive study programmes, based on the application of statistical procedures or model-based methods, has not been yet deeply developed. 
In contrast, despite the multi-cycle structure of the university system is effective since more a decade, the investigation appears to be limited to the computation of descriptive summaries in the national reports on the state of the higher education systems, that clearly does not allow for a constructive characterization of the transition phenomenon.
With regard to the Italian experience, the main reason lies in the lack of exhaustive and unified national archives of students' individual careers, that record the evolution and the outcomes of their academic paths. 

In the international literature, very few attempts have been implemented to go further basic descriptive analyses. 
For example, 
by means of a factor analysis \cite{Boer} explored the motivations that guide Bachelor graduates at universities of applied sciences to move to research-led universities for entering into a Master programme. 
\cite{Veroszta2013} and \cite{Nyusti2015}, instead, interpret the transition/exit of first level graduates as a selection mechanism and apply a logistic regression model to identify the role of several institutional factors on the continuation outcome in the Hungarian higher education context.
In \cite{Schomburg2011} a collection of studies from several European countries concerning Bachelor graduates is presented.
\cite{Cammelli2011} contribute to such a collection with the results of the most recent surveys on the Italian graduates carried out by the AlmaLaurea Inter-University Consortium, with a special focus on the international mobility, the employment condition and the study continuation decision of the Bachelor graduates. Although their analysis relies on rich and integrated data bases, it still remains at an exploratory level. 

In the attempt to go beyond a partial picture of the phenomenon and unveil the determinants of the transition dynamic, a multivariate setting that describes the causal relation between the outcome and a set of explanatory variables should be invoked.
Our work contributes to fill this gap in the Bachelor-Master transition literature by making use of a recent proposal of regression modeling for binary responses in the Bayesian QR framework,
that we review in the next section.


\section{Bayesian quantile regression modeling for binary responses}
\label{s:modeling}
In order to formalize the non-transition event, the response variable 
is 
represented by a binary indicator $y$, equal to 1 to denote the \textit{Bachelor-Master drop-out}, that is, the case of student's withdrawal after the achievement of the Bachelor degree, and equal to 0 in the case of continuation of the academic studies at Master level. 
In this work we consider the BQR approach in a Bayesian framework, for which
a continuous latent variable $y^*$
measuring the \textit{propensity} or \textit{willingness to drop-out} after the Bachelor qualification is introduced. 
The BQR model allows for studying the effect of known predictors on location measures of the latent propensity distribution other than the mean and can integrate the analysis
with the description of extreme behaviors and attitudes concerning the non-continuation decision.
In general, the application of the QR setup is also motivated as a more robust alternative to the traditional conditional mean model in the presence of outliers and heteroskedasticity \citep{Huber}. 
Moreover, the Bayesian paradigm considered here 
is particularly useful for modeling binary response variables since it solves some of the technical drawbacks faced when the frequentist approach is applied. As remarked by \cite{Benoit:Van}, the use of frequentist methods may exhibit difficulties in optimizing the regression parameters and in building confidence intervals for the estimates.
To our knowledge there is no previous application of the Bayesian BQR model for the analysis of the Bachelor-Master transition phenomenon.  

In this section we review the BQR model and detail how, from a Bayesian estimation perspective, the data augmentation method 
represents the key strategy to address inferential issues related to the discreteness of the outcomes.

\subsection{Model setup}
\label{ss:modset}
Let $\mathbf{y}=(y_1,\dots,y_n)$ be the vector of the observed binary outcomes, where the subscript $i=1,\dots,n$ indexes the sample units. 
For an arbitrary quantile level $\tau\in(0,1)$, the BQR model originally proposed by \cite{Manski1975} postulates that
\begin{align} 
\label{e:ytrunc} 
y_i&=I_{ [y^*_i\geq0]},\\
\label{e:quantreg}
y^*_i&=\mathbf{x}'_i\bm{\beta}_\tau+\epsilon_i,
\end{align} 
where $I_{[E]}$ denotes the indicator function of the event $E$, $\mathbf{x}_i=(1,x_{i1},\dots,x_{iK})'$ is the $(K+1)\times1$ vector of known predictors describing the individual profile with respect to the covariates, $\bm{\beta}_\tau=(\beta_{\tau0},\dots,\beta_{\tau K})'$ is the $(K+1)\times1$ vector of regression parameters depending on $\tau$ and $\epsilon_i$ is the random error component such that $\int_{-\infty}^0f_{\epsilon_i}(\epsilon_i)\,d\epsilon_i=\tau$.
Assumption \eqref{e:ytrunc} illustrates the latent variable interpretation of the BQR, that relies on the introduction of an underlying quantitative variable $y^*_i$ associated to each observed data point $y_i$.  
Equality \eqref{e:ytrunc}, in fact, expresses $y_i$ as the dichotomization of the corresponding latent continuous variable $y^*_i$.
As described by equality \eqref{e:quantreg}, an ordinary QR model is postulated on $y^*_i$, where the error variable $\epsilon_i$ is constrained to have the $\tau$-th quantile equal to zero.
The QR model  for quantitative responses were firstly introduced by \cite{Koenker:Bassett}. 
A general introduction on the topic can be found in \cite{Koenker2001}, whereas \cite{Yu2003} and \cite{Koenker2005} offer a review of the literature.
From \eqref{e:quantreg} it follows that 
$$Q_{y^*_i|\mathbf{x}_i,\bm{\beta}_\tau}(\tau)=\mathbf{x}'_i\bm{\beta}_\tau,$$
where $Q_{y^*_i|\mathbf{x}_i,\bm{\beta}_\tau}(\cdot)$ denotes the conditional quantile function of $y^*_i$. When $\tau=0.5$, equality \eqref{e:quantreg} becomes the popular conditional median regression model.
The theoretical justification of combining \eqref{e:ytrunc} and \eqref{e:quantreg} to construct a QR model for the analysis of binary outcomes is provided by the \textit{equivariance property} of quantile functions described in \cite{Powell1984}, stating that $Q_{g(y^*)}(\cdot)=g(Q_{y^*}(\cdot))$ when $g$ is a monotone nondecreasing function. 
In the BQR context this property applies since one has $g(\cdot)=I_{[\cdot]}$, leading to 
$$Q_{y_i|\mathbf{x}_i,\bm{\beta}_\tau}(\tau)=I_{\left[Q_{y_i^*|\mathbf{x}_i,\bm{\beta}_\tau}(\tau)\geq0\right]}=I_{[\mathbf{x}'_i\bm{\beta}_\tau\geq0]}.$$

\subsection{Model inference}
\label{ss:modest}
In the Bayesian domain, inference on BQR is efficiently solved by means of the \textit{data augmentation} strategy, that essentially allows to borrow in the discrete case the well-established estimation framework
of the QR for a continuous responses, as described in \cite{Benoit:Van} and \cite{Benoit2013}.
Auxiliary variables have been demonstrated to be convenient for both sample space completion and prior specification, since they facilitate the construction of sampling-based algorithms to conduct approximate inference \citep{Tanner:Wong}. 

Tracing back to the seminal work of \cite{Koenker:Bassett},
in the absence of parametric restrictions on the error terms, frequentist estimates $\hat{\bm{\beta}}_\tau$ for the QR model \eqref{e:quantreg} can be obtained as the solution of the following optimization problem
\begin{equation}
\label{e:optimum}
\hat{\bm{\beta}}_\tau=\underset{\bm{\beta}_\tau}{\operatorname{arg\,min}}\sum_{i=1}^n\rho_\tau(y^*_i-\mathbf{x}'_i\bm{\beta}_\tau),
\end{equation}
where $\rho_\tau(u)=u(\tau-I_{ [u<0]})=(|u|+(2\tau-1)u)/2$ is called \textit{loss} or \textit{check} \textit{function}.
\cite{Koenker:Machado} and \cite{Yu:Moyeed} pointed out that the resulting $\tau$-th \textit{regression quantile} $\hat{\bm{\beta}}_\tau$
coincides with the maximum likelihood estimate
under independent ALD for the unobserved error terms. 
This finding inspired a series of works on QR modeling in the Bayesian framework where, instead, the specification of the likelihood is needed. Recent contributions to the Bayesian literature can be found in \cite{Reed2009}, \cite{Li2010}, \cite{Kozumi2011}, \cite{Lum}, \cite{Alhamzawi2012}, \cite{Ji2012}, \cite{Alhamzawi:Yu} and \cite{Bernardi2015}.
%
To implement the Bayesian inference a convenient choice is the use of the exponential-gaussian mixture representation of the ALD, see \cite{Kotz}, \cite{Park:Casella} and \cite{Kozumi2011}. In particular, let $\epsilon\sim\text{ALD}(\mu,\sigma,\tau)$ with density function given by
\begin{equation*}
\label{e:ald}
f_{\text{ADL}}(\epsilon|\mu,\sigma,\tau)=\frac{\tau(1-\tau)}{\sigma}e^{-\rho_\tau\left(\frac{\epsilon-\mu}{\sigma}\right)}
\qquad \epsilon\in\mathbb{R},
\end{equation*}
where $\tau\in(0,1)$ is the skewness parameter, $-\infty<\mu<+\infty$ is the location parameter reflecting both the mode and the $\tau$-th quantile and $\sigma>0$ is the scale parameter. 
By following \cite{Kotz} and setting $\theta=\frac{1-2\tau}{\tau(1-\tau)}$ and $p^2=\frac{2}{\tau(1-\tau)}$, the random variable $\epsilon\sim\text{ALD}(0,1,\tau)$ admits the representation in terms of location-scale mixture of normals given by
\begin{equation}
\label{e:mixture}
\epsilon=\theta u+p\sqrt{u}\,z, 
\end{equation}
where $z\sim\text{N}(0,1)$ and $u\sim\text{Exp}(1)$ are mutually independent and N($\cdot$) and Exp($\cdot$) denote, respectively, the Gaussian and the Exponential distribution. 
The substitution of the equality \eqref{e:mixture} in \eqref{e:quantreg} leads to
\begin{equation}
\label{e:rewrite}
y^*_i=\mathbf{x}'_i\bm{\beta}_\tau+\theta u_i+p\sqrt{u_i}\,z_i, 
\end{equation}
implying that $y^*_i|u_i,\mathbf{x}_i,\bm{\beta}_\tau\sim\text{N}(\mathbf{x}'_i\bm{\beta}_\tau+\theta u_i,p^2u_i)$.
Thus, the mixture representation \eqref{e:rewrite} expands the likelihood specification into a favorable hierarchical structure, whose fundamental advantage is the possibility to transfer the normal linear model framework to the QR approach. 
As described by \cite{Reed2009} and \cite{Kozumi2011}, with the elicitation of conjugate normal priors on the regression parameters, the derivation of an efficient Gibbs sampling (GS) is straightforward. 
With the addition of equality \eqref{e:ytrunc},
the same inferential scheme can be easily extended into the Bayesian BQR estimation for the analysis of dichotomous responses \citep{Ji2012}.
In the discrete case, in fact, one can augment the observed binary data $\mathbf{y}$ with both the latent vectors $\mathbf{y}^*=(y_1^*,\dots,y_n^*)$ and $\mathbf{u}=(u_1,\dots,u_n)$ and write the 
complete-data likelihood as follows
\begin{equation*}
\label{e:complik}
\begin{split}
L_c(\bm{\beta}_\tau,\mathbf{y}^*,\mathbf{u})
&=\prod_{i=1}^n(y_iI_{[y^*_i\geq0]}+(1-y_i)I_{[y^*_i<0]})f(y^*_i|u_i,\mathbf{x}_i,\bm{\beta}_\tau)f(u_i|\mathbf{x}_i,\bm{\beta}_\tau).
\end{split}
\end{equation*}
%
The unique role played by the binary response is the truncation of the conditional normal density of $y_i^*$ 
on the suitable range. 
The resulting analytic expressions of the full-conditional distributions to be sampled in the GS 
are briefly recalled in the Appendix.
For the application presented in the next section, model estimation has been implemented in R \citep{Rsoft} with the recently released R package \texttt{bayesQR} developed by \cite{bayesQR}. 
For comparison purposes, in order to show the superiority of the BQR in capturing the Bachelor-Master drop-out features, we estimated also a more conventional Bayesian logistic regression model by employing the routines of the R package \texttt{BayesLogit} \citep{BayesLogit}. 

\section{Empirical application}
\label{s:empana}

In this paper we work with individual data regarding students who enrolled in a post-reform first level course in the School of Economics of Sapienza University of Rome and attained their Bachelor degree in one of the a.y. from 2009-2010 to 2012-2013. 
The data set was provided by the administrative offices of Sapienza University of Rome and collects several information on the four cohorts of undergraduates including their personal characteristics, educational background and academic career.
In this section we first describe the variables considered in the regression framework and the preliminary evidence emerging from the exploratory analysis and then discuss the results of the Bayesian BQR model.

\subsection{Definition of the variables}
\label{ss:defvar}

Here we are interested in 
identifying which factors impact on the decision of dropping-out of the university after the achievement of the 3-year first level qualification. 
For this purpose, we constructed a binary response variable $y_i$ for all $i=1,\dots,n$ as indicator of the non-continuation event. In more detail, we set $y_i=1$ if student $i$ is not enrolled in a new second level study programme at Sapienza within the two a.y. following the Bachelor degree attainment and $y_i=0$ otherwise. 
Thus, our main focus is on the Bachelor-Master drop-out of a specific academic institution, rather than on the abandonment of the entire university system. The case $y_i=1$, in fact, occurs when either the undergraduate decides to transfer to another institution or to leave the university system. The case $y_i=0$, instead, denotes the transition to a Master degree course and includes both the retention in the same school and the enrollment in second level programme of another school within Sapienza.
For each considered undergraduate cohort, the occurrence of the withdrawal event has been assessed in the two a.y. after the first level graduation.
Notice that \cite{Boer} considered the same time interval to define the transition decision in their analysis.
We argue the choice of the two-year period with the fact that the observed proportion of students who enroll in a second cycle after more than two a.y. from the completion of the the first cycle career is negligible. 
Moreover, this evidence agrees with the national statistics on the Bachelor-Master progression rate reported in \cite{ANVUR2014}, indicating that almost all the transitions occur without any break period between the two cycles. 

As potential predictors of the latent willingness to drop-out we considered variables concerning different aspects of the student's individual profile.
In line with the existing literature, our model specification includes personal characteristics, such as gender ($1=\text{Male}$, $0=\text{Female}$), age at first level graduation, citizenship ($1=\text{Italian}$, $0=\text{Other}$) and place of residence ($1=\text{Rome}$, $0=\text{Other}$). 
The administrative data provide also information on the family financial condition expressed by means of the Equivalent Economic Situation Indicator (ISEE).\footnote{The ISEE is a quantitative measure of the household economic status and represents the standard tool to request favorable conditions for the access to public services in Italy. In the academic context the ISEE is employed to apply personalized university fees, whose amount varies according to the contributory level of the student. 
The ISEE is equal to the ratio of the economic situation indicator, obtained by summing the incomes of each household member and 20\% the family heritage, and a coefficient that accounts for the number of members and specific characteristics of the family unit.}
The index is suitably rescaled to allow comparisons between family units that differ in size and composition. 
Similarly to \cite{Belloc2010} and \cite{Belloc2011}, in our analysis we categorized the ISEE values into four classes, respectively $0 \text{ \euro}\leq \text{ISEE}<10000\text{ \euro}$ (reference category), $10000\text{ \euro}\leq \text{ISEE}<20000\text{ \euro}$, $20000\text{ \euro}\leq \text{ISEE}<30000\text{ \euro}$ and $\text{ISEE}\geq 30000\text{ \euro}$. 
Unfortunately, further information on the household context are not collected by the administration. 
This prevented us to assess the effect of parental characteristics, such as the educational background and the occupational condition, although in previous works on university drop-out these variables were found to be associated with the non-continuation decision after the enrollment in the university, as discussed in \cite{Cipollone}, \cite{dHombres}, \cite{Dipietro-Cutillo} and \cite{Aina2013}. 

Regarding the student educational background, we considered a binary variable equal to 1 if the student granted the diploma in a classical or scientific lyceum and equal to 0 for other types of high secondary schools. 
With this dichotomization we stress the fundamental distinction between secondary schools that provide a more general and theoretical preparedness, such as the lyceums, and the technical or professional institutions, that are mainly oriented to a vocational training for the future exercise of specific professions. 
Besides the type of high school attended before the enrollment in the university, students were also asked by the administration to specify their high school leaving mark, that in the present analysis has been rescaled in the interval [0.6, 1].

The remaining predictors, instead, concern student's academic career. 
Administrative data contain several variables describing the individual study path and, specifically, detailed information on the type of first level qualification earned by the student and on the academic performance.  
Regarding the first aspect, we stress that the School of Economics at Sapienza University of Rome offers a wide range of 3-year degree courses. 
By exploiting the grouping of the degree courses with similar learning objectives into homogeneous 
degree classes, as prescribed by the MD 509/99 and subsequently maintained by the MD 270/04, the 3-year degree courses of the School of Economics are formally categorized into two main classes: Economics and Business Management. We adopted this classification to characterize the field of study through a binary variable such that $1=\text{Economics}$ and $0=\text{Business Management}$. 
Moreover, since the new MD 270/04 came into effect in the a.y. 2008-2009 and impacted to a substantial 
extent on the students' study plans in the School of Economics, as previously discussed in Section \ref{s:reform} we decided to account also for the set of academic regulations that the student underwent with an indicator variable defined as $1=\text{MD 270/04}$ and $0=\text{MD 509/99}$.

The role of the academic performance on the choice to leave the university is widely debated in the drop-out literature, although it seems to depend on the specific context, as demonstrated by the heterogeneous evidence collected so far. 
With the achievement of the Bachelor degree, the final graduation mark could be regarded as the candidate measure of student's performance. Indeed, 
the final grade is obtained from the combination of two main components: the average mark of passed exams and the actual duration of the studies. 
The former variable involves the credits described in Section~\ref{s:reform} and it is equal to the weighted average of the exam grades with weights given by the corresponding amount of credits. 
The latter variable is, instead, related to an aspect of great concern for the Italian academic institutions, known as \textit{out-of-course phenomenon}.
Formally, a student is classified by the administration as \textit{out-of-course graduate} when he/she attains the degree qualification in a time greater than the legal duration of the study course. 
Unlike other European university systems, where the students have to pass the exams within the prescribed period (or at most with a short delay) in order to successfully conclude their studies, in Italy they can continue to be enrolled in the university and graduate after the scheduled period 
as long as they pay the due fees. 
Thus, the out-of-course condition can be coded with a dummy variable equal to 1 for out-of-course undergraduates and to 0 for students who earned the Bachelor degree within the scheduled duration.
In order to better capture different aspects of student's ability, we preferred to work with the weighted average mark  (WAM) of passed exams and the binary indicator of the out-of-course graduation, rather than with the single final degree mark. In this way we can discriminate how the level of acquired knowledge and the fulfillment of the scheduled time limit separately act on the latent response. 
Additionally, we considered an interaction term between the WAM and the out-of-course condition to account for the major difficulty of exhibiting a good academic performance and graduating on time rather than after a longer period of study. 

Finally, we examined also the possible presence of a time trend, or ``cohort effect'', by involving in the linear predictor a categorical variable with four levels indicating the a.y. of graduation, where the a.y. 2009-2010 is assumed as the reference category.

\subsection{Exploratory analysis}
\label{ss:exanalysis}

\begin{table}[h]
\centering
\begin{threeparttable}
\caption{Summary statistics.} 
\addtolength{\tabcolsep}{+15pt}    
\begin{tabular}
{lcc}
Categorical variable & \% & \\
\hline
Bachelor-Master drop-out & & \\
\quad No & 72.2  & \\
\quad Yes & 27.8 & \\
Gender & & \\
\quad Female & 54.0 & \\
\quad Male & 46.0 & \\
Academic year of graduation & & \\
\quad 2009-2010 & 20.8 & \\
\quad 2010-2011 & 25.7 & \\
\quad 2011-2012 & 27.0 & \\
\quad 2012-2013 & 26.5 & \\
Citizenship & & \\
\quad Italian & 95.4 & \\
\quad Other & 4.6 & \\
Place of residence & & \\
\quad Rome & 70.8 & \\
\quad Other & 29.2 & \\
ISEE & & \\
\quad [0,10000) & 18.4 & \\
\quad [10000,20000) & 26.9 & \\
\quad [20000,30000) & 20.5 & \\
\quad $\geq$30000 & 34.3 & \\
Degree class & & \\
\quad Business Management & 86.9 & \\
\quad Economics & 13.1 & \\
MD & & \\
\quad 509/99 & 26.6 & \\
\quad 270/04 & 73.4 & \\
Lyceum & & \\
\quad No & 39.5 & \\
\quad Yes & 60.5 & \\
Out-of-course & & \\
\quad No & 31.7 & \\
\quad Yes & 68.3& \\
\hline
Quantitative variable & mean & s.d.\\
\hline
Age at graduation & 24.1 & 3.53  \\
WAM & 24.2 & 1.86  \\
High school diploma mark & 0.8 & 0.12  \\
\hline
 \end{tabular}
\label{t:sumstat}
\begin{tablenotes}
      \small
      \item $n=2655$ observations
      \item ISEE $=$ Equivalent Economic Situation Indicator
      \item WAM $=$ weighted average mark of passed exams
      \item MD $=$ Ministerial Decree
      \item Source: administrative offices of Sapienza University of Rome
\end{tablenotes}
\end{threeparttable}
\end{table}
\begin{table}[h]
\centering
\begin{threeparttable}
\caption{Bachelor-Master drop-out rates (\%) by specific characteristics for all the Bachelor graduates (Overall) and by cohort.} 
\addtolength{\tabcolsep}{-3pt}    
\begin{tabular}
{lccccc}
Variable & Overall & 2009-2010 & 2010-2011 & 2011-2012 & 2012-2013 \\
\hline
Academic year of graduation & & & & &\\
\quad 2009-2010 & 21.6 & - & - & - & -\\
\quad 2010-2011 & 26.1 & - & - & - & -\\
\quad 2011-2012 & 29.9 & - & - & - & -\\
\quad 2012-2013 & 32.0 & - & - & - & -\\
Gender & & & & &\\
\quad Female & 25.3 & 20.1 & 20.5 & 28.4 & 31.4\\
\quad Male & 30.7 & 23.7 & 32.4 & 31.9 & 32.6\\
Citizenship & & & & &\\
\quad Italian & 27.5 & 20.8 & 25.9 & 29.4 & 32.3\\
\quad Other & 33.9 & 42.9 & 29.4 & 41.9 & 25.7\\
Place of residence & & & & &\\
\quad Rome & 27.6 & 21.2 & 24.9 & 30.7 & 31.9\\
\quad Other & 28.2 & 22.6 & 28.8 & 28.0 & 32.4\\
ISEE & & & & &\\
\quad [0,10000) & 26.6 & 21.1 & 25.0 & 30.3 & 28.9\\
\quad [10000,20000) & 25.8 & 19.7 & 24.6 & 28.4 & 29.7 \\
\quad [20000,30000) & 25.4 & 19.8 & 23.4 & 25.0 & 33.1 \\
\quad $\geq$30000 & 31.3 & 25.2 & 29.4 & 33.6 & 34.6 \\
Degree class & & & & &\\
\quad Business Management & 26.5 & 18.8 & 26.1 & 28.5 & 30.4\\
\quad Economics & 36.2 & 35.5 & 25.6 & 40.2 & 43.9\\
MD & & & & &\\
\quad 509/99 & 34.7 & 25.4 & 37.6 & 40.0 & 72.7\\
\quad 270/04 & 25.3 & 14.9 & 21.2 & 28.3 & 28.5\\
Lyceum & & & & &\\
\quad No & 31.1 & 19.7 & 34.1 & 32.3 & 37.9 \\
\quad Yes & 25.6 & 23.2 & 20.6 & 28.6 & 28.6 \\
Out-of-course & & & & &\\
\quad No & 18.6 & 14.8 & 19.7 & 20.5 & 19.5 \\
\quad Yes & 32.0 & 25.6 & 29.5 & 33.7 & 36.8 \\
\hline
 \end{tabular}
\label{t:sumstatby}
\begin{tablenotes}
      \small
      \item $n=2655$ observations
      \item ISEE $=$ Equivalent Economic Situation Indicator
      \item WAM $=$ weighted average mark of passed exams
      \item MD $=$ Ministerial Decree
      \item Source: administrative offices of Sapienza University of Rome
\end{tablenotes}
\end{threeparttable}
\end{table}
\begin{figure}[t]
\centering
\includegraphics[scale=.45]{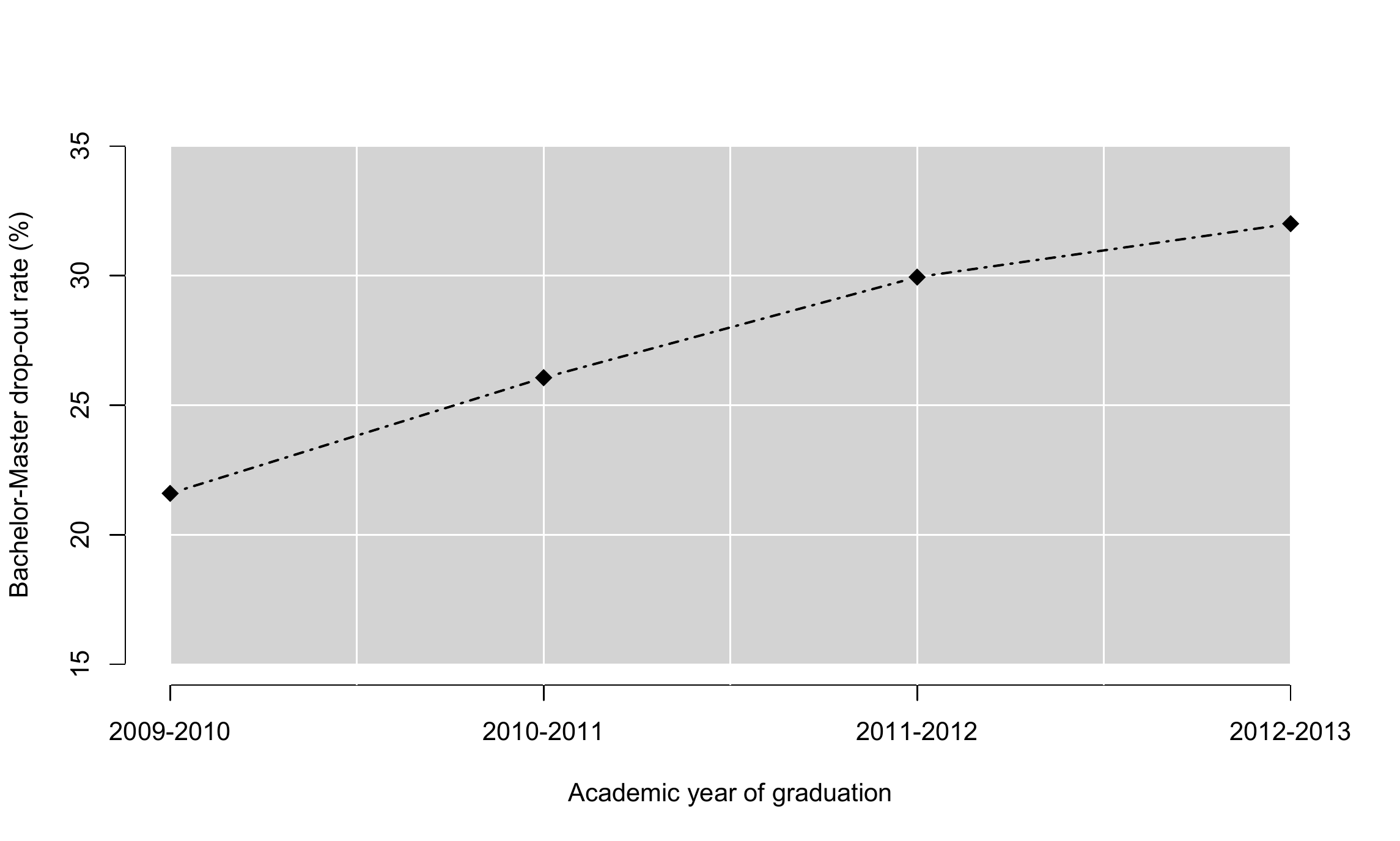}
\caption{Bachelor-Master drop-out rate (\%) by academic year of graduation.}
\label{fig:TimeSeries}
\end{figure}
\begin{figure}[b]
\centering
\includegraphics[scale=.45]{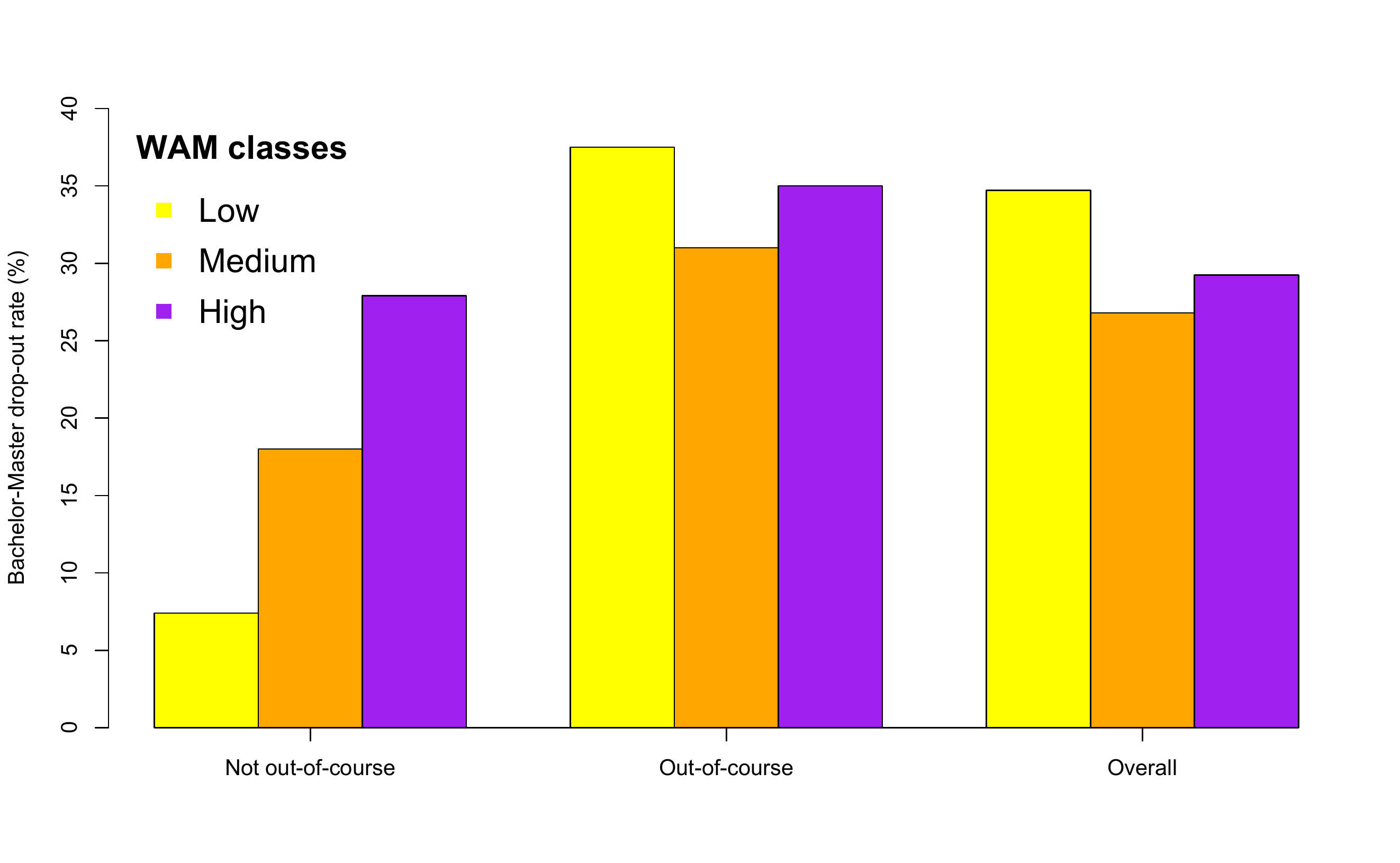}
\caption{Bachelor-Master drop-out rate (\%) by WAM classes: overall percentage (right) and the one stratified by out-of-course condition (left and center).}
\label{fig:MosaicDropvsWAM}
\end{figure}
\begin{figure}[t]
\centering
\includegraphics[scale=.45]{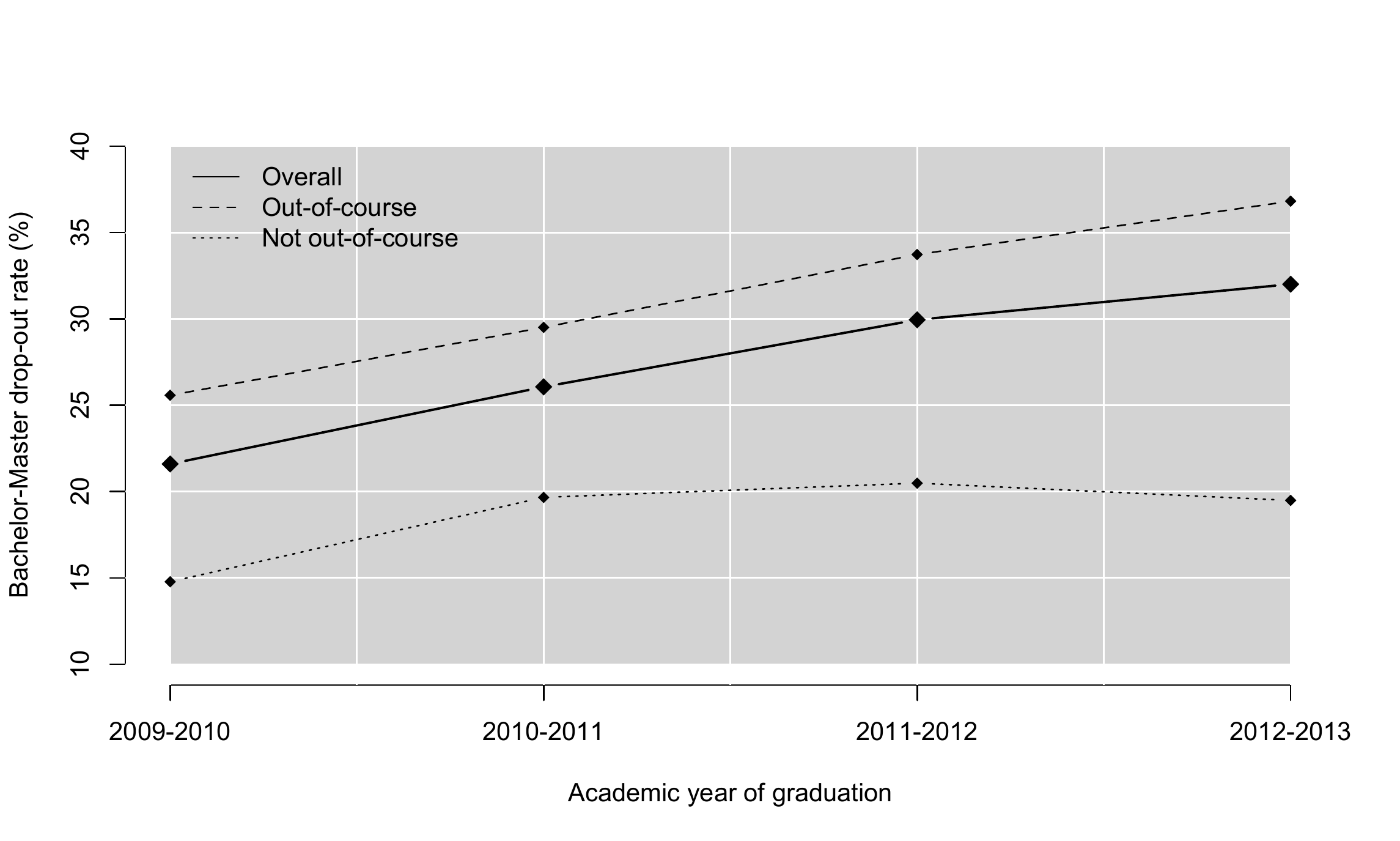}
\caption{Bachelor-Master drop-out rate (\%) by academic year of graduation (solid line) and by out-of-course condition (dashed lines).}
\label{fig:NotInTime}
\end{figure}
\begin{figure}[t]
\centering
\subfloat{ 
\includegraphics[scale=.39]{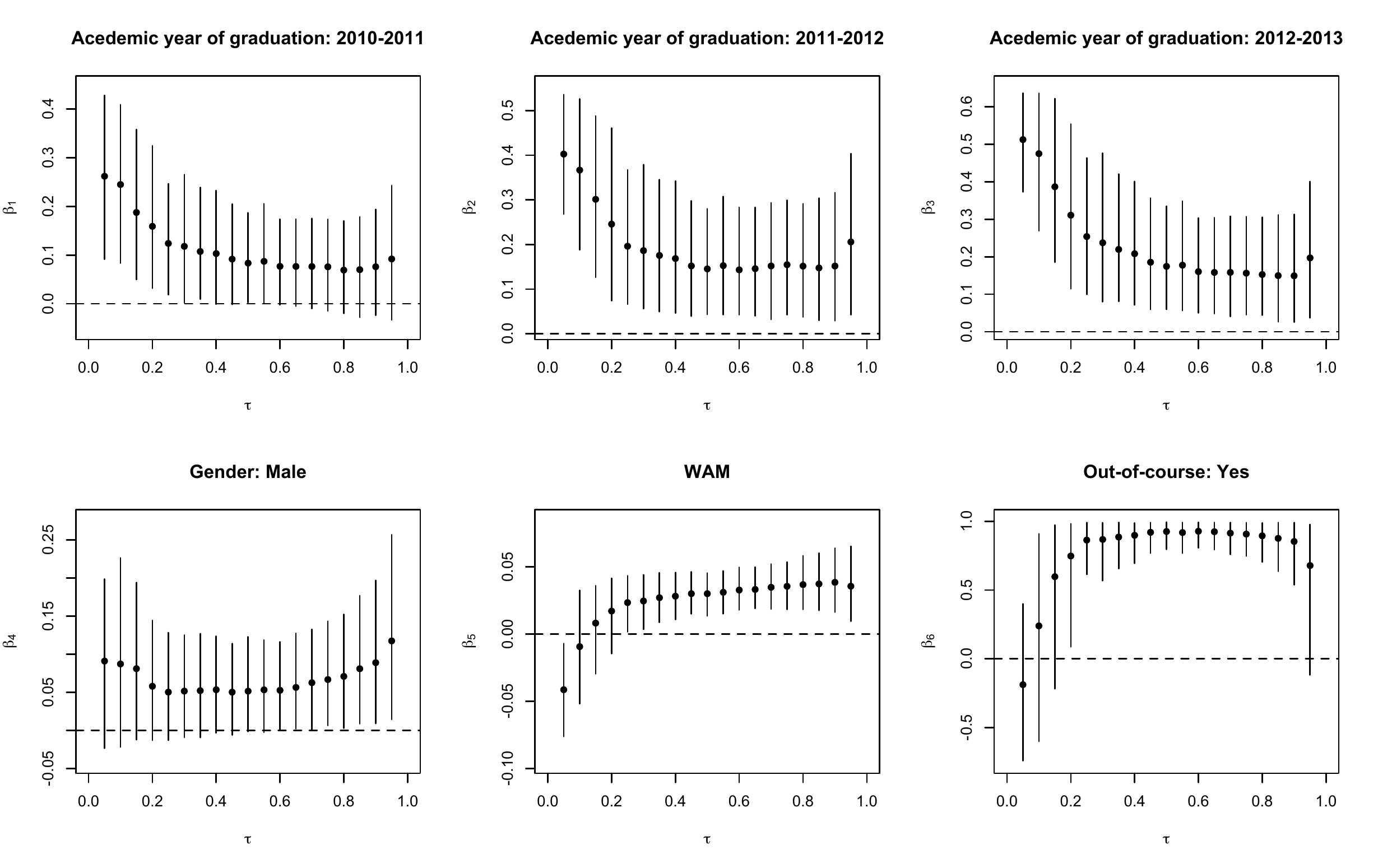}
\label{fig:Forest1}}\\
\subfloat{ 
\includegraphics[scale=.39]{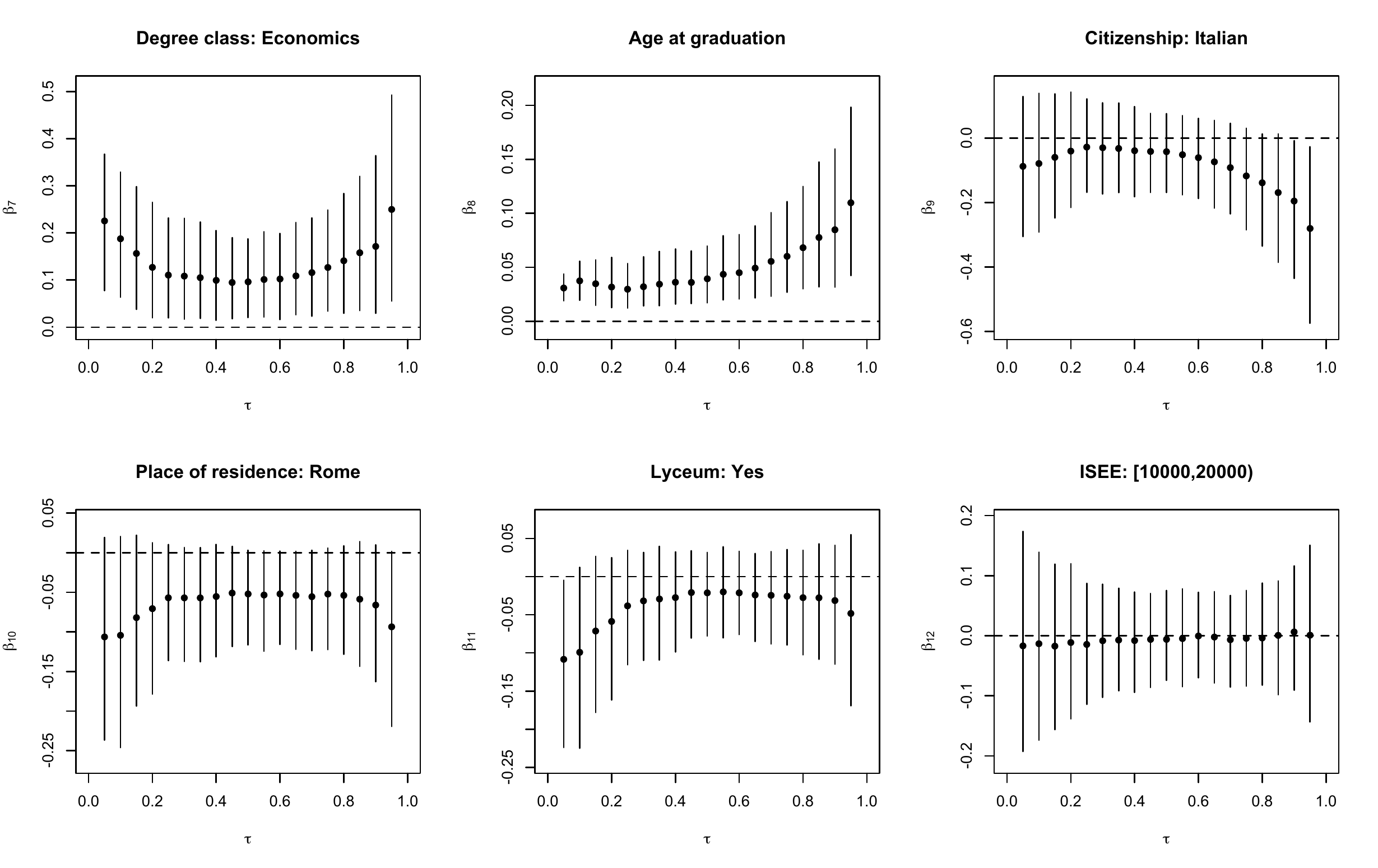}
\label{fig:Forest2}}\\
\subfloat{ 
\includegraphics[scale=.39]{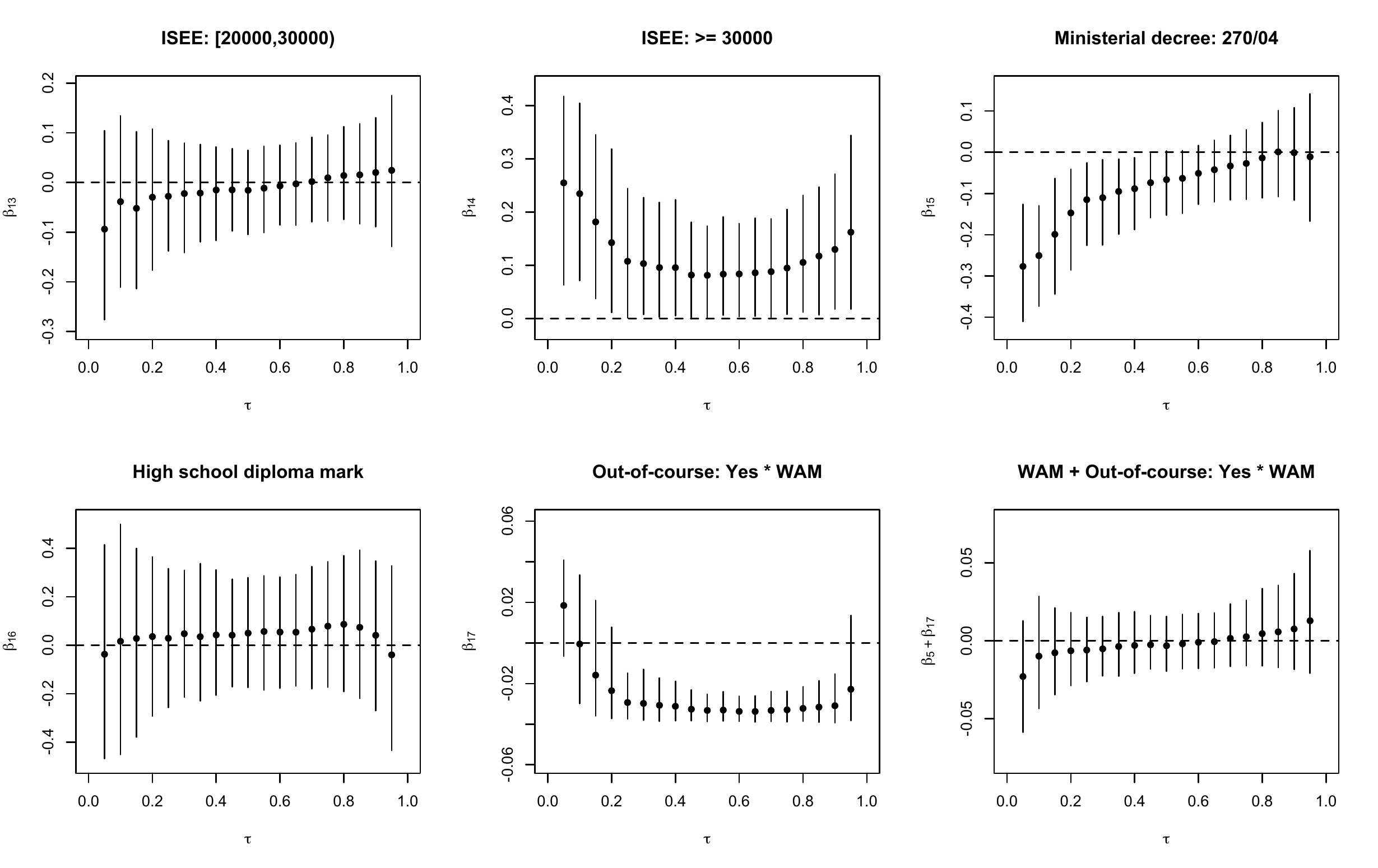}
\label{fig:Forest3}}
\caption{Forest plots of the posterior estimates of the BQR parameters for selected values of the quantile level $\tau$. Points indicate the posterior means and segments represent the corresponding 95\% HPQ credible intervals.}
\label{fig:forest}
\end{figure}

The data set provided by the administrative offices of Sapienza University of Rome contains a total of  $n=2655$ individual profiles relative to four cohorts of Bachelor graduates.
As shown by the summary statistics reported in Table~\ref{t:sumstat}, there is a slight majority of females (54.0\%) among undergraduates, the most part (70.8\%) of them is resident in Rome and only 4.6\% is a non-Italian citizen. 
Regarding the educational background, 60.5\% undergraduates attended a lyceum (classic or scientific) and the rescaled average high school leaving mark
is 0.8. 

In the considered four-year period, the Business Management study field largely prevails on the Bachelor degree in Economics, with 86.9\% students who achieved the qualification in this degree class. Moreover, 73.4\% Bachelor graduates underwent the academic regulations established by the MD 270/04 during their first level career. 
The mean age at first level graduation is 24.1 years and less than one third (31.7\%) of the students completed the first cycle within the legal duration of the degree programme. These values on the average age at graduation and the regularity of the first cycle studies are in line with the national indicators reported in \cite{ANVUR2014}.
%

As far as the response variable is concerned, the overall Bachelor-Master transition rate observed at the School of Economics in the considered period is 72.2\%, that turns out to be comparable with the progression percentages estimated at national level for the Economics-Statistics area (Figure~\ref{fig:TimeSeriesTransition}).
Only few cases of school change within Sapienza have been observed, indicating a high correspondence of the second cycle studies with the first level career. 
An analogous tendency for the Economics-Statistics area of study is highlighted in \cite{Cammelli2011}, showing that a massive percentage of Bachelor graduates in this field of study confirms their first cycle studies by progressing at Master level in the same school. 

Consequently, in the considered period more than one forth (27.8\%) of the undergraduates decided to retire from Sapienza after the achievement of the Bachelor degree. 
The interest in the non-continuation decision is reinforced by the fact that the School of Economics experienced a manifest increment of the withdrawal phenomenon during the reference period, with a 10-points growth of the Bachelor-Master drop-out proportion 
from 21.6\% for the cohort 2009-2010 up to 32.0\% for the cohort 2012-2013, as represented in Figure~\ref{fig:TimeSeries}. 

Table~\ref{t:sumstatby} shows the non-continuation proportions by specific undergraduates' characteristics, employed as preliminary tool to explore the association with the covariates of interest.
A first notable aspect to be stressed is that the drop-out decision appears to be strongly associated with the out-of-course condition, since students who graduated after the scheduled end of the course more frequently choose to interrupt their educational path in Sapienza after the Bachelor degree than those who regularly conclude the first cycle. This evidence is remarked by the large gap between the dashed lines in Figure~\ref{fig:NotInTime}, pointing out that this differential pattern characterizes all the four Bachelor cohorts and, indeed, the difference between out-of-course and regular undergraduates visibly increases over time.


Interestingly, a remarkable differential evidence emerges also for the variables describing the type of first level qualification. 
The non-continuation percentage is considerably higher for the Bachelor graduates in Economics than for those who achieved the degree in Business Management (36.2\% for the former versus 26.5\% for the latter).
Additionally, students subject to the MD 509/99 regulations during their first level career exhibit a greater Bachelor-Master drop-out rate (34.7\%) than students undergone the MD 270/04 (25.3\%). 


To facilitate the preliminary analysis of the relation between the continuous WAM variable and the binary outcome, we transformed the WAM of passed exams into an ordered factor
with three levels reflecting, respectively, a low, medium and high academic performance and then computed the drop-out percentage within each class. The barplot on the right in Figure~\ref{fig:MosaicDropvsWAM}  (Overall) indicates higher withdrawal rates for the extreme WAM classes and, hence, the presence of a marginal non-monotonic relation with the outcome. Interestingly, by conditioning on the out-of-course condition, a similar pattern is observed also for the out-of-course undergraduates, whereas a clear positive influence of the WAM is revealed for regular Bachelors (not out-of-course students). This finding further supports the possible usefulness of an interaction term between the two variables in the regression equation to gain a more in-depth understanding of the academic performance as potential determinant of the non-continuation event. 

Regarding the personal characteristics and the educational background, Table~\ref{t:sumstatby} points out 
a greater propensity to interrupt the studies among males, foreign students and among those who do not come from a lyceum.

By exploring also the association between the response variable and the household economic condition, 
one can observe that  
the first three ISEE classes have a comparable drop-out rate around 26\%, whereas the non-continuation percentage among undergraduates whose ISEE is over 30.000 \text{\euro} is higher (31.3\%).

Of course, all the above preliminary findings provide only a crude description of the interruption decision. 
To suitably enquire into the effect of the explanatory variables on the withdrawal event, a multiple regression analysis is needed. 

\subsection{Results from the Bayesian BQR analysis}
\label{ss:qranalysis}
   
For the analysis of the Bachelor-Master transition we estimated a series of Bayesian BQR models by varying the quantile level $\tau$ in the regular grid $\{0.05,0.10,\dots,0.90,0.95\}$. 
Considering a burn-in phase of 1000 iterations, further 10000 GS drawings have been sampled from the posterior distribution to derive the final estimates. Algorithm convergence was assisted by the routines in the R package \texttt{coda} \citep{Plummer} and, specifically, by checking the mixing of the trace-plots and the extent of the sample autocorrelation for each parameter. 

The inferential results of the BQR can be conveniently summarized by means of the forest plots constructed on the normalized coefficient estimates of each explanatory variable, as displayed in Figure~\ref{fig:forest}.
As recall by \cite{Kordas2006}, in fact, the regression parameter vector $\bm{\beta}_\tau$ in the BQR model is identified up to a scale, 
see also \cite{Manski1985} for an in-depth explanation and treatment of the identifiability issue. Thus, 
in order to allow for comparisons among different conditional quantiles, 
\cite{Kordas2006} suggests to unify the scale by normalizing the slope parameters with the Euclidean norm $||\cdot||$.
By denoting with $\bm{\beta}_\tau^{-c}$ the regression parameter vector deprived of the intercept $\beta_{\tau0}$, we considered posterior inference on $\bm{\beta}_\tau^{-c}$/$||\bm{\beta}_\tau^{-c}||$.
For each specific predictor, the forest plot simultaneously describes the point and interval estimate of the corresponding normalized regression coefficient as a function of the quantile level $\tau$ (Figure~\ref{fig:forest}).
Specifically, the points in the forest plots represent the posterior means, whereas the segments illustrate the 95\% Highest Posterior Density (HPD) credible intervals derived from the MCMC samples. The dashed horizontal line at zero was added to facilitate the identification of significant effects on the latent quantitative response $y^*$. 

The significant results for the degree class in Economics over all the considered quantile levels confirm an important differential continuation behavior between the two 
study courses, 
that is, Bachelor graduates in Economics have a higher probability to conclude their study experience after the first cycle than 
undergraduates in Business Management.
One possible interpretation could be the presence of competitive and appealing Master degree programmes for the Economics area of study in other universities, that could 
lead students to change institution.  

Regarding the academic performance, firstly we can note that the regularity of the study path is related with the continuation decision. Specifically, students who do not complete their studies on time have a higher probability to retire from the institution, as indicated by the interval estimates largely above the dashed line in the corresponding plot of Figure~\ref{fig:forest}. 
Moreover, the addition of the interaction term in the BQR model provides a very enlightening interpretation on the impact of the WAM on the binary outcome, suggesting a different effect of the performance variable for out-of-course and regular students. 
The significant and positive estimates of the coefficient associated to the WAM regressor
reveal that, for almost all quantile levels, this variable influences the interruption choice of regular students (reference group). 
This means that, among students who fulfill the regular duration of the programme, those who conclude their first level studies with a higher WAM have a greater probability of dropping out after the Bachelor degree. 
Several conjectures can be formulated about the estimated effect of the student's performance. A greater Bachelor-Master drop-out rate for larger WAM values could be explained with the fact that brilliant students have a higher chance to find a job after the first cycle studies or could be even employed before concluding their 3-year degree course. 
On the other hand, best-performing students take their decisions with a more careful and critical evaluation; thus, if they realize that other institutions offer a more training programme matching their expectations, they are more prone to move.
By summing the WAM coefficient for the reference group with the interaction estimate, one obtains the effect of the WAM for out-of-course undergraduates, displayed in the last plot of Figure~\ref{fig:forest}. The latter regression coefficient turns out to be not statistically significant over the full range of $\tau$ values, meaning that the level of acquired competence does not modify the drop-out decision of out-of-course students. 
In the absence of the interaction term, this important difference on the relevance of the academic performance in the two groups of students would have been missed. 

Unlike previous studies emphasizing the role of the type of diploma and of the high school leaving mark, in our analysis these variables do not emerge as relevant predictors of the progression at Master level. It is possible, in fact, that the educational background exerts an effect on the continuation decision in the first phase of the university career, that in general immediately follows the high school experience, but not on the transition to the second cycle studies.

Regression coefficients associated to the cohort indicators clearly show the growing trend of Bachelor-Master drop-out rate in the considered period and, in particular, a significant increase for the 
last two Bachelor cohorts.
 
The coefficient for the age at graduation exhibits a significant positive value across all quantiles with a manifest growing trend. This means that age at graduation not only contributes to explain the drop-out decision, but its relevance is even stronger for students with a high propensity to leave the university after the first cycle. The positive effect of age could be argued with the fact that older students could desire to not further delay their entry in the labour market or could be already employed and, hence, less motivated to put effort into a new degree course.


The forest plot in Figure~\ref{fig:forest} regarding the estimates for the MD indicator is the one that better exemplifies the usefulness of the BQR approach.
The MD 270/04 is estimated to have a negative effect on the entire latent non-continuation propensity, but its magnitude and significance considerably vary across quantiles.
By moving from higher to lower quantiles, one has that the estimates associated to the MD 270/04 remarkably decrease and vary from not statistically significant values to significant effects, specifically on the quantiles below the median. Moreover, the influence of the MD 270/04 becomes more and more pronounced on the left tail of the underlying distribution. 
This implies that students with lower propensity to withdraw are sensible to the type of academic regulations when they have to choose whether to continue at Master level or not and, in particular, those who experienced the MD 270/04 are less likely to drop-out after the first level qualification.
In order to show the usefulness of the BQR approach, we performed a Bayesian logistic regression model with the same regressors. By comparing the results obtained from the two methods (Table~\ref{t:postmean}), it is possible to point out that the difference between the two MD is not captured by the latter model, since the corresponding effect 
is estimated to be not significant. 
These findings stress the importance of having more information on the whole distribution of the phenomena rather than on the mean value only.
To a reduced extent, a similar monotone behavior of the posterior estimates can be also highlighted for the effect of the gender. Being male affects positively the drop-out probability but its impact is significant only for quantile levels above 0.75. This suggests that the difference between males and females mainly concerns students with high propensity to interrupt. 
The Italian citizenship is estimated to have a negative effect that becomes stronger over higher quantiles. However, the corresponding coefficient is significant only on the 90-th and 95-th percentile. This borderline evidence could be due to very low prevalence of foreign students in the present study.
In accordance with the preliminary analysis, the BQR approach does not indicate a difference in the continuation behavior for students belonging to the first three ISEE classes, whereas undergraduates with a household economic indicator at least equal to 30000 \text{\euro} have a significant higher probability to drop-out. 
Finally, the place of residence turns out to be not significant in the description of the students' continuation choice for the present study, although the negative sign is consistent with several drop-out analyses stating that being resident in the city where the university is located facilitates the continuation of the academic studies.

\begin{table}[h]
\centering
\begin{threeparttable}
\caption{Posterior estimates of the Bayesian logistic regression model fitted to the data set of the Bachelor graduates at the School of Economics. Point and interval estimates are respectively the posterior means and the 95\% HPD credible intervals.} 
\addtolength{\tabcolsep}{-3pt}    
\begin{tabular}
{lrrr}
 & Posterior mean & 95\% HPD Lower & 95\% HPD Upper\\
\hline

Intercept   & -9.081 & -11.750 & -6.357$^{*}$ \\
Academic year of graduation: 2010-2011   &   0.355 & 0.048 & 0.643$^{*}$ \\
Academic year of graduation: 2011-2012   &  0.620 & 0.301 & 0.929$^{*}$ \\
Academic year of graduation: 2012-2013   &  0.711 & 0.378 & 1.029$^{*}$ \\
Gender: Male   &  0.221 & 0.037 & 0.418$^{*}$ \\
Out-of-course: Yes   & 4.565 & 1.736 & 7.515$^{*}$ \\
Degree class: Economics   &  0.429 & 0.175 & 0.680$^{*}$ \\
Age at graduation   & 0.157 & 0.122 & 0.191$^{*}$ \\
Citizenship: Italian  & -0.265 & -0.714 & 0.186 \\
Place of Residence: Rome & -0.204 & -0.401 & 0.003 \\
Lyceum: Yes & -0.094 & -0.298 & 0.110 \\
ISEE: [10000,20000) &  -0.017 & -0.291 & 0.273 \\
ISEE: [20000,30000) & -0.020 & -0.337 & 0.280 \\
ISEE: $\geq$30000 &  0.356 & 0.085 & 0.643$^{*}$ \\
MD: 270/04  & -0.274 & -0.524 & 0.000 \\
High school diploma mark  &  0.190 & -0.667 & 1.109 \\
WAM   & 0.150 & 0.053 & 0.255$^{*}$ \\
Out-of-course: Yes $\times$ WAM & -0.168 & -0.287 & -0.054$^{*}$ \\
WAM $+$ Out-of-course: Yes $\times$ WAM & -0.017 & -0.087 & 0.057 \\
\hline
 \end{tabular}
\label{t:postmean}
\begin{tablenotes}
      \small
      \item $n=2655$ observations
      \item ISEE $=$ Equivalent Economic Situation Indicator
      \item WAM $=$ weighted average mark of passed exams
      \item MD $=$ Ministerial Decree
      \item Number of Burn-in iterations $ = $ 1000
      \item Number of GS iterations $ = $ 10000
      \item * $ = $ 95\% HPD interval not including zero
\end{tablenotes}
\end{threeparttable}
\end{table}

\section{Concluding remarks and future developments}
\label{s:conc}

The reforms of the Bologna process have transformed the academic training in a stepwise path, whose multi-cycle structure is the core feature.
Although a wide literature concerns the problem of university drop-out referred to the very first years after the enrollment, very little has been explored with respect to the transition from the first to the second cycle academic studies. 
In this paper we described a BRQ model from the Bayesian estimation perspective as effective device to describe the withdrawal event after the first level qualification. We approached the failed Bachelor-Master transition as a drop-out problem and tried to understand the main determinants of the non-continuation choice by exploiting the QR model setup. This method permits to highlight different student behaviors by means of a latent propensity scale measuring the willingness to drop-out.
From a Bayesian point of view, 
the combination of a data augmentation step with the location-scale mixture representation of the ALD allows to address
inference in a straightforward manner, without relying on asymptotic properties and computational demanding methods required, instead, under the frequentist approach. Moreover, the Bayesian framework allows for an easy derivation of the posterior credible intervals providing well understandable measures of the statistical uncertainty associated to the estimates.

From an empirical point of view, one of the main contributions of this work concerns the assessment of a critical aspect of the Italian university context, that is, the out-of-course phenomenon, and the identification of the significant benefits of graduating on time on the continuation decision at Master level. 
Consequently, some factual interventions to reduce the out-of-course rate
should be planned, such as a more accurate revision of the course programmes and of the corresponding credits and/or a different organization of the exam sessions during the a.y..
Interestingly, our analysis also points out that the changes introduced by the MD 270/04 are associated with a reduction of the non-continuation rate after the Bachelor degree, that instead does not emerge from the traditional logistic regression model. 
This result provides an important feedback at institutional level, since it concretely suggests to the relevant authorities how to improve the academic regulations and plan the future interventions.
Obviously, the MD variable defined for the present Italian case can be suitably reinterpreted in the other national contexts to assess the impact of the academic regulations in the other countries.




A natural extension of the present Bachelor-Master drop-out analysis could be the application of the methods to the data of the entire university, as well as to the regional or national ones. 
In this regard, a generalization of the Bayesian BQR for handling the multilevel structure of the individual observations should be developed, due to the typical nested organization of the universities where degree courses are nested into departments which are in turn nested into schools.
Some advantages of the multilevel modeling include the possibility to capture the heterogeneity among different degree courses or departments and to derive relative performance measures for comparisons purposes. 
Moreover, the multilevel approach allows to introduce course- or department-level covariates and to assess their impact on the students' decisions. 
This remarkably broadens the scope of policy makers and can guide a more efficiently allocation of the resources aimed at enhancing university effectiveness. A contribution of multilevel analysis involving university drop-out can be found in \cite{Meggiolaro2013}.
 
In conclusion, as emphasized in \cite{ANVUR2014}, the relation between the Bachelor and the Master degree 
certainly needs further investigation, in order to clarify to what extent the first and the second cycle studies have to be regarded as intermediate stages of a unique training path rather than as separated steps of the academic experience.


\appendix
\section*{Appendix}
\subsubsection*{Full-conditionals determination for the Bayesian binary quantile regression model}
By adopting $\bm{\beta}_\tau\sim\text{N}_{K+1}(\bm{\beta}_0,\mathbf{B}_0)$ as prior distribution for the BQR parameters, 
thanks to the equation \eqref{e:rewrite}
the full conditional of the latent $y_i^*$ for all $i=1,\dots,n$ is
\begin{equation*}
\label{e:fullystar1}
\begin{split}
y^*_i|y_i,u_i,\mathbf{x}_i,\bm{\beta}_\tau\sim\begin{cases}
\text{N}(\mathbf{x}'_i\bm{\beta}_\tau+\theta u_i,p^2u_i)I_{[y^*_i\geq0]} & \quad\text{if } y_i=1, \\
\text{N}(\mathbf{x}'_i\bm{\beta}_\tau+\theta u_i,p^2u_i)I_{[y^*_i<0]} & \quad\text{otherwise}.
\end{cases}
\end{split}
\end{equation*}
Generating from a truncated normal distribution can be accomplished by means of the algorithms proposed by \cite{Geweke}.
By indicating with $\pi(u_i|y^*_i,y_i,\mathbf{x}_i,\bm{\beta}_\tau)$ the full conditional of $u_i$, we can note that the presence of $y_i$ in the conditioning right-hand side 
is redundant once $y^*_i$ is known. Thus, 
\begin{equation*}
\label{e:fullystar1}
\begin{split}
\pi(u_i|y^*_i,y_i,\mathbf{x}_i,\bm{\beta}_\tau)&=\pi(u_i|y^*_i,\mathbf{x}_i,\bm{\beta}_\tau)
\propto f(y^*_i|u_i,\mathbf{x}_i,\bm{\beta}_\tau)f(u_i|\mathbf{x}_i,\bm{\beta}_\tau)\\
&\propto u_i^{-1/2}e^{-\frac{1}{2}\bigl(\frac{(y^*_i-\mathbf{x}'_i\bm{\beta}_\tau)^2}{p^2u_i}+\frac{2+\theta^2}{p^2}u_i\bigr)},
\end{split}
\end{equation*}
where one can recognize the kernel of a Generalized Inverse Gaussian (GIG) distribution given by 
\begin{equation*}
u_i|y^*_i,y_i,\mathbf{x}_i,\bm{\beta}_\tau\sim\text{GIG}(1/2,(y^*_i-\mathbf{x}'_i\bm{\beta}_\tau)^2/p^2,2+\theta^2/p^2).
\end{equation*}
%
Finally, thanks to the conjugacy of the normal prior density for the regression parameters with the Gaussian distribution of the latent responses $\mathbf{y}^*$ conditionally on $\mathbf{u}$, the full conditional of $\bm{\beta}_\tau$ is still a member of the multivariate normal family, obtained as 
\begin{equation*}
\bm{\beta}_\tau|\mathbf{y}^*,\mathbf{y},\mathbf{u},\mathbf{X}\sim\text{N}_{K+1}(\hat{\bm{\beta}}_\tau,\hat{\mathbf{B}}_\tau),
\end{equation*}
where the hyperparameters $\bm{\beta}_0$ and $\mathbf{B}_0$ are updated as in the traditional Bayesian normal linear regression model according to the following expressions
\begin{equation*}
\hat{\mathbf{B}}^{-1}_\tau=\tau^{-2}\mathbf{X}'\mathbf{U}^{-1}\mathbf{X}+\mathbf{B}^{-1}_0\qquad\text{and}\qquad
\hat{\bm{\beta}}_\tau=\hat{\mathbf{B}}_\tau\biggl(\tau^{-2}\mathbf{X}'\mathbf{U}^{-1}(\mathbf{y}^*-\theta \mathbf{u})+\mathbf{B}^{-1}_0\bm{\beta}_0\biggr)
\end{equation*}
with $\mathbf{U}=\text{diag}(\mathbf{u})$.

\bibliographystyle{apa}
\bibliography{BiblioQuantile}

\begin{thebibliography}{}

\bibitem[\protect\astroncite{Aina}{2013}]{Aina2013}
Aina, C. (2013).
\newblock Parental background and university dropout in {I}taly.
\newblock {\em Higher Education}, 65(4):437--456.

\bibitem[\protect\astroncite{Alhamzawi and Yu}{2013}]{Alhamzawi:Yu}
Alhamzawi, R. and Yu, K. (2013).
\newblock Conjugate priors and variable selection for {B}ayesian quantile
  regression.
\newblock {\em Computational Statistics \& Data Analysis}, 64:209--219.

\bibitem[\protect\astroncite{Alhamzawi et~al.}{2012}]{Alhamzawi2012}
Alhamzawi, R., Yu, K., and Benoit, D.~F. (2012).
\newblock {B}ayesian adaptive {L}asso quantile regression.
\newblock {\em Statistical Modelling}, 12(3):279--297.

\bibitem[\protect\astroncite{ANVUR}{2014}]{ANVUR2014}
ANVUR (2014).
\newblock {\em Rapporto sullo stato del sistema universitario e della ricerca}.
\newblock Agenzia Nazionale di Valutazione del Sistema Universitario e della
  Ricerca.

\bibitem[\protect\astroncite{Ballarino and Bratti}{2009}]{Ballarino}
Ballarino, G. and Bratti, M. (2009).
\newblock Field of study and university graduates' early employment outcomes in
  {I}taly during 1995--2004.
\newblock {\em Labour}, 23(3):421--457.

\bibitem[\protect\astroncite{Belloc et~al.}{2010}]{Belloc2010}
Belloc, F., Maruotti, A., and Petrella, L. (2010).
\newblock University drop-out: an italian experience.
\newblock {\em Higher Education}, 60(2):127--138.

\bibitem[\protect\astroncite{Belloc et~al.}{2011}]{Belloc2011}
Belloc, F., Maruotti, A., and Petrella, L. (2011).
\newblock How individual characteristics affect university students drop-out: a
  semiparametric mixed-effects model for an {I}talian case study.
\newblock {\em Journal of Applied Statistics}, 38(10):2225--2239.

\bibitem[\protect\astroncite{Benoit et~al.}{2013}]{Benoit2013}
Benoit, D.~F., Alhamzawi, R., and Yu, K. (2013).
\newblock {B}ayesian {L}asso binary quantile regression.
\newblock {\em Computational Statistics}, 28(6):2861--2873.

\bibitem[\protect\astroncite{Benoit et~al.}{2014}]{bayesQR}
Benoit, D.~F., Alhamzawi, R., Yu, K., and Van~den Poel, D. (2014).
\newblock bayes{QR}: {B}ayesian quantile regression.
\newblock {\em R package version 2.2}.

\bibitem[\protect\astroncite{Benoit and Van~den Poel}{2012}]{Benoit:Van}
Benoit, D.~F. and Van~den Poel, D. (2012).
\newblock Binary quantile regression: a {B}ayesian approach based on the
  asymmetric laplace distribution.
\newblock {\em Journal of Applied Econometrics}, 27(7):1174--1188.

\bibitem[\protect\astroncite{Bernardi et~al.}{2015}]{Bernardi2015}
Bernardi, M., Gayraud, G., and Petrella, L. (2015).
\newblock Bayesian tail risk interdependence using quantile regression.
\newblock {\em Bayesian Analysis}, 10(3):553--603.

\bibitem[\protect\astroncite{Biggeri et~al.}{2001}]{Biggeri}
Biggeri, L., Bini, M., and Grilli, L. (2001).
\newblock The transition from university to work: a multilevel approach to the
  analysis of the time to obtain the first job.
\newblock {\em Journal of the Royal Statistical Society: Series A (Statistics
  in Society)}, 164(2):293--305.

\bibitem[\protect\astroncite{Cammelli et~al.}{2011}]{Cammelli2011}
Cammelli, A., Antonelli, G., Di~Francia, A., Gasperoni, G., and Sgarzi, M.
  (2011).
\newblock Mixed outcomes of the {B}ologna process in {I}taly.
\newblock In {\em Employability and mobility of {B}achelor graduates in
  Europe}, pages 143--170. Springer.

\bibitem[\protect\astroncite{Cipollone and Cingano}{2007}]{Cipollone}
Cipollone, P. and Cingano, F. (2007).
\newblock University drop-out-the case of italy.
\newblock {\em Bank of Italy Temi di Discussione (Working Paper) No}, 626.

\bibitem[\protect\astroncite{de~Boer et~al.}{2010}]{Boer}
de~Boer, H., Kolster, R., and Vossensteyn, H. (2010).
\newblock Motives underlying {B}achelors--{M}asters transitions: The case of
  {D}utch degree stackers.
\newblock {\em Higher Education Policy}, 23(3):381--396.

\bibitem[\protect\astroncite{d'Hombres}{2007}]{dHombres}
d'Hombres, B. (2007).
\newblock The impact of university reforms on dropout rates and students?
  status: Evidence from {I}taly.
\newblock Technical report, JRC European Commission.

\bibitem[\protect\astroncite{Di~Pietro and Cutillo}{2008}]{Dipietro-Cutillo}
Di~Pietro, G. and Cutillo, A. (2008).
\newblock Degree flexibility and university drop-out: The {I}talian experience.
\newblock {\em Economics of Education Review}, 27(5):546--555.

\bibitem[\protect\astroncite{Geweke}{1991}]{Geweke}
Geweke, J. (1991).
\newblock Efficient simulation from the multivariate normal and {S}tudent-t
  distributions subject to linear constraints and the evaluation of constraint
  probabilities.
\newblock In {\em Computing {S}cience and {S}tatistics: Proceedings of the 23rd
  {S}ymposium on the {I}nterface}, pages 571--578.

\bibitem[\protect\astroncite{Grilli}{2006}]{Grilli2006}
Grilli, L. (2006).
\newblock Multilevel models for the analysis of the transition from university
  to work.
\newblock In {\em VIII International Meeting on Quantitative Methods for
  Applied Sciences}.

\bibitem[\protect\astroncite{Huber}{1981}]{Huber}
Huber, P.~J. (1981).
\newblock {\em Robust {S}tatistics}.
\newblock Wiley New York.

\bibitem[\protect\astroncite{Ji et~al.}{2012}]{Ji2012}
Ji, Y., Lin, N., and Zhang, B. (2012).
\newblock Model selection in binary and {T}obit quantile regression using the
  {G}ibbs sampler.
\newblock {\em Computational Statistics \& Data Analysis}, 56(4):827--839.

\bibitem[\protect\astroncite{Koenker}{2005}]{Koenker2005}
Koenker, R. (2005).
\newblock {\em Quantile regression}.
\newblock Number~38. Cambridge University Press.

\bibitem[\protect\astroncite{Koenker and Bassett}{1978}]{Koenker:Bassett}
Koenker, R. and Bassett, G.~J. (1978).
\newblock Regression quantiles.
\newblock {\em Econometrica: Journal of the Econometric Society}, pages 33--50.

\bibitem[\protect\astroncite{Koenker and Hallock}{2001}]{Koenker2001}
Koenker, R. and Hallock, K. (2001).
\newblock Quantile regression: An introduction.
\newblock {\em Journal of Economic Perspectives}, 15(4):43--56.

\bibitem[\protect\astroncite{Koenker and Machado}{1999}]{Koenker:Machado}
Koenker, R. and Machado, J.~A. (1999).
\newblock Goodness of fit and related inference processes for quantile
  regression.
\newblock {\em Journal of the American Statistical Association},
  94(448):1296--1310.

\bibitem[\protect\astroncite{Kordas}{2006}]{Kordas2006}
Kordas, G. (2006).
\newblock Smoothed binary regression quantiles.
\newblock {\em Journal of Applied Econometrics}, 21(3):387--407.

\bibitem[\protect\astroncite{Kottas and Gelfand}{2001}]{Kottas2001}
Kottas, A. and Gelfand, A.~E. (2001).
\newblock Bayesian semiparametric median regression modeling.
\newblock {\em Journal of the American Statistical Association},
  96(456):1458--1468.

\bibitem[\protect\astroncite{Kotz et~al.}{2001}]{Kotz}
Kotz, S., Kozubowski, T.~J., and Podg{\'o}rski, K. (2001).
\newblock {\em The {L}aplace distribution and generalizations: A revisit with
  applications to communications, economics, engineering, and finance}.
\newblock Birkh\"auser Boston, Inc., Boston, MA.

\bibitem[\protect\astroncite{Kozumi and Kobayashi}{2011}]{Kozumi2011}
Kozumi, H. and Kobayashi, G. (2011).
\newblock Gibbs sampling methods for {B}ayesian quantile regression.
\newblock {\em Journal of Statistical Computation and Simulation},
  81(11):1565--1578.

\bibitem[\protect\astroncite{Larsen et~al.}{2013}]{Larsen2013a}
Larsen, M.~S., Kornbeck, K.~P., Kristensen, R.~M., Larsen, M.~R., and
  Sommersel, H.~B. (2013).
\newblock Dropout phenomena at universities: What is dropout? {W}hy does
  dropout occur? {W}hat can be done by the universities to prevent or reduce
  it? {A} systematic review.
\newblock Technical report, Department of Education, Aarhus University.

\bibitem[\protect\astroncite{Li et~al.}{2010}]{Li2010}
Li, Q., Xi, R., Lin, N., et~al. (2010).
\newblock {B}ayesian regularized quantile regression.
\newblock {\em Bayesian Analysis}, 5(3):533--556.

\bibitem[\protect\astroncite{Lum et~al.}{2012}]{Lum}
Lum, K., Gelfand, A.~E., et~al. (2012).
\newblock Spatial quantile multiple regression using the {A}symmetric {L}aplace
  {P}rocess.
\newblock {\em Bayesian Analysis}, 7(2):235--258.

\bibitem[\protect\astroncite{Manski}{1975}]{Manski1975}
Manski, C.~F. (1975).
\newblock Maximum score estimation of the stochastic utility model of choice.
\newblock {\em Journal of Econometrics}, 3(3):205--228.

\bibitem[\protect\astroncite{Manski}{1985}]{Manski1985}
Manski, C.~F. (1985).
\newblock Semiparametric analysis of discrete response: Asymptotic properties
  of the maximum score estimator.
\newblock {\em Journal of Econometrics}, 27(3):313--333.

\bibitem[\protect\astroncite{Meggiolaro et~al.}{2013}]{Meggiolaro2013}
Meggiolaro, S., Giraldo, A., and Clerici, R. (2013).
\newblock A multilevel competing risks model for analysis of university
  students' careers: evidence from {I}taly.
\newblock Technical report, Working Paper Series 9, University of Padua,
  Department of Statistical Sciences.

\bibitem[\protect\astroncite{Ny{\"u}sti et~al.}{2015}]{Nyusti2015}
Ny{\"u}sti, S., Veroszta, V., et~al. (2015).
\newblock Institutional effects on {B}achelor-{M}aster-level transition.
\newblock {\em International Journal of Social Sciences}, 4(1):39--61.

\bibitem[\protect\astroncite{Park and Casella}{2008}]{Park:Casella}
Park, T. and Casella, G. (2008).
\newblock The {B}ayesian {L}asso.
\newblock {\em Journal of the American Statistical Association},
  103(482):681--686.

\bibitem[\protect\astroncite{Plummer et~al.}{2006}]{Plummer}
Plummer, M., Best, N., Cowles, K., and Vines, K. (2006).
\newblock {CODA}: Convergence diagnosis and output analysis for {MCMC}.
\newblock {\em R news}, 6(1):7--11.

\bibitem[\protect\astroncite{Powell}{1984}]{Powell1984}
Powell, J.~L. (1984).
\newblock Least absolute deviations estimation for the censored regression
  model.
\newblock {\em Journal of Econometrics}, 25(3):303--325.

\bibitem[\protect\astroncite{Pozzoli}{2009}]{Pozzoli}
Pozzoli, D. (2009).
\newblock The transition to work for {I}talian university graduates.
\newblock {\em Labour}, 23(1):131--169.

\bibitem[\protect\astroncite{{R Core Team}}{2014}]{Rsoft}
{R Core Team} (2014).
\newblock {\em R: A Language and Environment for Statistical Computing}.
\newblock R Foundation for Statistical Computing, Vienna, Austria.

\bibitem[\protect\astroncite{Reed and Yu}{2009}]{Reed2009}
Reed, C. and Yu, K. (2009).
\newblock A {P}artially {C}ollapsed {G}ibbs sampler for {B}ayesian quantile
  regression.
\newblock Technical report, Department of Mathematical Sciences, Brunel
  University.

\bibitem[\protect\astroncite{Schomburg and Teichler}{2011}]{Schomburg2011}
Schomburg, H. and Teichler, U. (2011).
\newblock {\em Employability and mobility of {B}achelor graduates in Europe}.
\newblock Springer.

\bibitem[\protect\astroncite{Smith and Naylor}{2001}]{Smith-Naylor}
Smith, J.~P. and Naylor, R.~A. (2001).
\newblock Dropping out of university: a statistical analysis of the probability
  of withdrawal for {UK} university students.
\newblock {\em Journal of the Royal Statistical Society: Series A (Statistics
  in Society)}, 164(2):389--405.

\bibitem[\protect\astroncite{Tanner and Wong}{1987}]{Tanner:Wong}
Tanner, M.~A. and Wong, W.~H. (1987).
\newblock The calculation of posterior distributions by data augmentation.
\newblock {\em Journal of the American Statistical Association},
  82(398):528--540.

\bibitem[\protect\astroncite{Veroszta}{2013}]{Veroszta2013}
Veroszta, Z. (2013).
\newblock {T}he way to master programmes -- an examination of the selection
  mechanisms in the bachelor/master transition in higher education.
\newblock In {\em {H}ungarian {G}raduates 2011}, pages 9--35. Educatio Public
  Services Non-profit LLC, Department of Higher Education.

\bibitem[\protect\astroncite{Windle et~al.}{2013}]{BayesLogit}
Windle, J., Polson, N.~G., and Scott, J.~G. (2013).
\newblock Bayes{L}ogit: {B}ayesian logistic regression.
\newblock {\em R Package Version 0.2-4}.

\bibitem[\protect\astroncite{Yu et~al.}{2003}]{Yu2003}
Yu, K., Lu, Z., and Stander, J. (2003).
\newblock Quantile regression: applications and current research areas.
\newblock {\em Journal of the Royal Statistical Society: Series D (The
  Statistician)}, 52(3):331--350.

\bibitem[\protect\astroncite{Yu and Moyeed}{2001}]{Yu:Moyeed}
Yu, K. and Moyeed, R.~A. (2001).
\newblock {B}ayesian quantile regression.
\newblock {\em Statistics \& Probability Letters}, 54(4):437--447.

\end{thebibliography}

\end{document}